\documentclass[preprint,prb,amsmath,amssymb,preprintnumbers]{revtex4}
\usepackage{graphicx}% Include figure files
\usepackage{dcolumn}% Align table columns on decimal point
\usepackage{bm}% bold math
\usepackage[bookmarks,hypertex]{hyperref}
\usepackage{multirow}
\usepackage{rotating}
\usepackage[utf8]{inputenc}

\begin{document}

\title{Elastic properties and electronic structures of antiperovskite-type InNCo$_3$ and InNNi$_3$}

\author{Z. F. Hou}
\affiliation{Department of Physics, Fudan University, Shanghai, 200433, P. R. China}

%\date{\today}
%\pacs{71.15.Ap; 71.15.Mb; 71.20.Lp; 78.20.Ke}

\begin{abstract}
% Text of abstract
We have performed the first-principles calculations to study the
elasticity, electronic structure, and magnetism of InNCo$_3$ and
InNNi$_3$. The independent elastic constants are derived from the
second derivative of total energy as a function of strain, and the
elastic modulus are predicted according to the Voigt-Reuss-Hill
approximation. Our calculations show that the bulk modulus of InNCo$_3$ is slightly larger than that of InNNi$_3$ due to a smaller lattice constant for InNCo$_3$. For InNCo$_3$ the ferromagnetic state is energetically preferable to the paramagnetic state,
while the ground state of InNNi$_3$ is a stable paramagnetic (non-magnetic) state. This is due to the different strength of 2\textit{p}-3\textit{d} hybridization for the N-Co and N-Ni atoms in InNCo$_3$ and InNNi$_3$. 

\end{abstract}

%\begin{keyword}
%Elastic properties \sep Electronic structures \sep InNCo$_3$ \sep
%First-principles calculations
%\PACS 31.15.A- \sep 71.15.Mb \sep 74.25.Jb \sep 74.25.Ld
%\end{keyword}

%\end{frontmatter}

\maketitle

%%%%% main text %%%%%
\section{\label{sec:intro}Introduction}

The ternary nitrides or carbides with the general formula AXM$_3$ (
A: divalent or trivalent element; X: carbon or nitrogen; and M:
transition metal) are already known for several decades
~\cite{Goodenough70,Chern92,Jager93}. These compounds crystalize in
a cubic anti-perovskite structure (A: cube-corner position; X:
body-center position; M: face-center position) and exhibit a wide
range of interesting physical properties~\cite{Goodenough70}, such
as giant magneto-resistance~\cite{Kim2001} and nearly zero
temperature coefficient resistivity~\cite{Chi2001}. They have
renewedly attracted considerable attention due to the discovery of
superconductivity at $\sim$8 K in intermetallic compound
MgCNi$_3$~\cite{He01}.

Considering the Ni-rich composition, it is expected that the
ferromagnetism could exist in MgCNi$_3$. However, the absence of
ferromagnetism was observed in experiment~\cite{He01} for MgCNi$_3$. From the
electronic structures obtained by the first-principles
calculations~\cite{Singh01,Shim01,Rosner02,Wu2009251}, the
non-ferromagnetic ground state of MgCNi$_3$ is ascribed to a reduced
Stoner factor that results from a strong hybridization between the
Ni-3$d$ and C-2$p$ electrons. For other Ni-based ternary carbides
ACNi$_3$ (e.g., A = Al, Ca, In, Zn, and Cd), the first-principles
calculations~\cite{Wu2009251,Okoye05,Sieberer07,Zhong07,Wu20084232}
also show that the ground states of these compounds are
non-magnetic and the C-Ni bonding exhibits nearly same
characteristics as the one in MgCNi$_3$. Therefore, this indicates
that the change of composition A  could not induce the
ferromagnetism in ACNi$_3$. On the other side, it raises a question
whether the change of composition X or M can lead to the appearance
of ferromagnetism in AXM$_3$ or not.

Very recently, the antiperovskite-type compounds InN$_y$Co$_3$ and
InN$_y$Ni$_3$ ($y\sim$ 1.0 and 0.8, respectively) have been
synthesized by solid-gas reactions of metal powders with NH$_3$ and
they have been reported to have spin-glass-like properties based on
the measurements of temperature dependence
magnetization.~\cite{Cao093353} The recent first-principles
calculations~\cite{Sieberer07} showed that the non-stoichiometry
could affect the magnetic properties of ACNi$_3$ (e.g., AlCNi$_3$
and GaCNi$_3$) and suggested that the tendencies toward magnetism
found in experiments~\cite{Dong05,Tong06Al,Tong06Ga,Tong07} for
these compounds should be explained by the deviation of the Ni/C
atomic ratio from the ideal stoichiometry. To shed more light on the
understanding on the magnetic properties of InNCo$_3$ and InNNi$_3$
reported in experiment~\cite{Cao093353}, it is of great importance
to theoretically study the electronic structures of these two compounds
as well as the nature of the N-Co and N-Ni bondings.

In order to completely understand the electronic structures and
magnetic ground states of InNCo$_3$ and
InNNi$_3$ with cubic anti-perovskite structure, we carried out the first-principles calculations on these
compounds using the pseudopotentials method with plane-wave basis
set within the local density approximation and the generalized
gradient approximation. Since the elastic properties of a solid are highly associated with various fundamental solid-state properties such as phonon spectra, specific heat, Debye temperature, and so on, we have calculated the independent elastic constant and the elastic moduli of InNCo$_3$ and InNNi$_3$.

\section{\label{sec:method}Computational details}

All calculations on antiperovskite-type InNCo$_3$ and InNNi$_3$ were
performed using the Quantum ESPRESSO code~\cite{Baroni}, which is
based on the density functional theory (DFT)~\cite{Kohn65}. The
electronic exchange-correlation potential was calculated within the
local density approximation (LDA)~\cite{Ceperley80,Perdew81} and the
generalized gradient approximation using the scheme of
Perdew-Burke-Ernzerhof (PBE)~\cite{Perdew96}. The spin polarization
was also considered in the calculation in order to assess the
magnetic properties of these compounds. Electron-ion interaction was
represented by the norm-conserving optimized~\cite{Rappe90} designed
nonlocal pseudopotentials. The 4\emph{d} electrons are explicitly
included in the valence of In. The electronic wavefunctions were
expanded by the plane waves up to a kinetic energy cutoff of 55 Ry.
The \textbf{k}-point sampling in Brillouin zone (BZ) of simple cubic
lattice was treated with the Monkhorst-Pack scheme
\cite{Monkhorst76} and a 20$\times$20$\times$20 \textbf{k}-point
mesh (i.e., 286 irreducible points in the first BZ) was used. The
chosen plane-wave cutoff and number of \textbf{k} points were
carefully checked to ensure that the total energy was converged to
be better than 1 mRy/cell. The total energies are obtained as a
function of volume and they are fitted with the Birch-Murnaghan
3rd-order equation of states (EoS)~\cite{Birch47} to give the
equilibrium lattice constant and other ground state properties.
During the calculation of density of states (DOS), a dense
\textbf{k}-point mesh of 30$\times$30$\times$30 is used, the total
DOS is computed by the tetrahedron method~\cite{Lehmann77}, and the
atomic-projected DOS is calculated by the L\"{o}wdin
populations~\cite{Portal95}.

For a cubic crystal, its independent elastic constants are $c_{11}$,
$c_{12}$, and $c_{44}$. To determine the elastic constants of
InNCo$_3$ and InNNi$_3$ by means of the curvature of the internal
energy versus the strain curves~\cite{Mehl93,Wu07425}, three strain
modes~\cite{Hou081651} are adopted and their nonzero strains are as
follows: (1) $\epsilon_{11}=\epsilon_{22}=\delta$,
$\epsilon_{33}=(1+\delta)^{-2}-1$; (2)
$\epsilon_{11}=\epsilon_{22}=\epsilon_{33}=\delta$; and (3)
$\epsilon_{12}=\epsilon_{21}=\delta/2$,
$\epsilon_{33}=\delta^2/(4-\delta^2)$. The deformation magnitudes
$\delta$ from -0.012 to 0.012 in the step of 0.03 are applied in the
first and second strain modes, and $\delta$ from -0.04 to 0.04 in
the step 0.01 are adopted in the third strain mode. Once the
independent elastic constants for single crystal properties are
obtained through the above procedure, the elastic moduli (e.g., the
shear modulus and the bulk modulus) of polycrystalline aggregates
can be estimated according to the Voigt-Reuss-Hill
approximation~\cite{Voigt28,Reuss29,Hill52}. In the Voigt
average~\cite{Voigt28}, the shear modulus and the bulk modulus of
cubic lattice are given by
\begin{equation}\label{eq:1}
G_V=\frac{1}{5}\left[(c_{11}-c_{12})+3c_{44}\right]
\end{equation}
and
\begin{equation}\label{eq:2}
B_V= \frac{1}{3}(c_{11}+2c_{12}),
\end{equation}
while in the Reuss average~\cite{Reuss29} they are given by
\begin{equation}\label{eq:3}
G_R=\frac{5}{4(s_{11}-s_{12})+3s_{44}}
\end{equation}
and
\begin{equation}\label{eq:4}
B_R= \frac{1}{3s_{11}+6s_{12}}
\end{equation}
with the relations $c_{44}=s_{44}^{-1}$,
$c_{11}-c_{12}=(s_{11}-s_{12})^{-1}$, and
$c_{11}+2c_{12}=(s_{11}+2s_{12})^{-1}$ in the cubic lattice, where
$s_{ij}$ are the elastic compliance constants. Therefore, $G_R$ and
$B_R$ in the cubic lattice can be rewritten as
\begin{equation}\label{eq:5}
G_R=\left[\frac{4}{5}(c_{11}-c_{12})^{-1}+\frac{3}{5}c_{44}^{-1}\right]^{-1},
\end{equation}
and
\begin{eqnarray}\label{eq:6}
\nonumber
B_R&=&\frac{1}{3}\left[(c_{11}+2c_{12})\right]\\
   &=&B_V.
\end{eqnarray}
In the Hill empirical average~\cite{Hill52}, the shear modulus and
the bulk modulus are taken as $G=\frac{1}{2}(G_V+G_R)$ and
$B=\frac{1}{2}(B_V+B_R)$, respectively. Knowing $G$ and $B$, the
Young's modulus $E$ and Poisson's ratio $\nu$, which are frequently
measured for polycrystalline materials when investigating their
hardness, can be calculated from the isotropic relations:
\begin{equation}\label{eq:7}
E=\frac{9BG}{3B+G}
\end{equation}
and
\begin{equation}\label{eq:8}
\nu=\frac{3B-2G}{2(3B+G)} .
\end{equation}

\section{Results and Discussions}

\subsection{\label{sec:struct}Structural properties}
In experiment with the powder X-ray diffraction patterns, Cao et
al~\cite{Cao093353} have reported that InN$_y$Co$_3$ and
InN$_y$Ni$_3$ ($y\sim$ 1.0 and 0.8, respectively) have the cubic
anti-perovskite structure with the space group 221($Pm\bar{3}m$) and
the corresponding lattice parameters were 3.854 \AA~and 3.844 \AA,
respectively. Starting from the experimental data, we have
calculated the total energies of unit cell at a series of volumes
for each compound in the paramagnetic (PM) and ferromagnetic (FM)
states. The results are presented in Fig.~\ref{fig:1}. It is found
that the energy difference between the FM and PM states is -0.0397
eV (-0.226 eV) in the LDA (GGA) calculations for InNCo$_3$ and 0.0
eV for InNNi$_3$. The total magnetic moment of InNCo$_3$ is about
2.14 $\mu_B$ (2.91 $\mu_B$) and the local magnetic moment of each In ion is about 0.69 $\mu_B$ (0.94 $\mu_B$) in LDA (GGA) calculations. The total magnetic moment of InNNi$_3$ and the local magnetic moment of each Ni atom are zero. These indicate that
the ferromagnetic state is energetically favorable to InNCo$_3$ and the ground state of InNNi$_3$ is paramagnetic state
(non-magnetic). The obtained equilibrium lattice constant
($a_0$), bulk modulus ($B$), and first pressure derivative of bulk
modulus ($B^{\prime}$) of InNCo$_3$ and InNNi$_3$ are listed in
Table~\ref{tab:1}. In our calculations, the predicted lattice constant of InNCo$_3$ is slightly larger than that of InNNi$_3$, which is oppsite to the trend reported in experiment~\cite{Cao093353}. This may be due to the deviation of the Ni/N atomic ratio from the ideal stoichiometry in experiment for InN$_y$Ni$_3$. In addition, it can be seen that the deviations of the LDA (GGA)
lattice constants of both the InNCo$_3$ and InNNi$_3$ with respect
to the experimental values are less than 2.6\% (1.0\%). That is to
say, the calculated equilibrium lattice constants of InNCo$_3$ and
InNNi$_3$ are in excellent agreement with the experimental
data~\cite{Cao093353}.

\subsection{\label{sec:poly}Elastic properties}

The calculated independent elastic constants for single crystal of
InNCo$_3$ and InNNi$_3$ are listed in Table~\ref{tab:2}. Based on
the Voigt-Reuss-Hill approximation~\cite{Voigt28,Reuss29,Hill52},
the elastic moduli of InNCo$_3$ and InNNi$_3$ are estimated and the
results are listed in Table~\ref{tab:2}. For the bulk moduli of
InNCo$_3$ and InNNi$_3$, the estimations based on the independent
elastic constants agree well those obtained by the fit of the
Birch-Murnaghan 3rd-order EoS. To the best of our knowledge, no
experimental data or theoretical results for the elasticity of
InNCo$_3$ and InNNi$_3$ compounds have been reported up to now.
Considering that the elastic properties of ZnNNi$_3$, InNSc$_3$, and InCNi$_3$ with cubic
anti-perovskite structure have been studied recently and the
theoretical results are available in
literature~\cite{Maurizio09,Wu20084232,lichong09,Shein10}, it will
be meaningful to compare them with those of InNCo$_3$ and InNNi$_3$
compounds. The order of bulk moduli of these five compounds from low to high is:
\textit{B}(InNSc$_3$) $<$ \textit{B}(InNNi$_3$) $<$
\textit{B}(InCNi$_3$) $<$ \textit{B}(InNCo$_3$) $<$
\textit{B}(ZnNNi$_3$). This could be understood from the trend in
lattice constants ($a$) of these compounds (i.e., $a$(InNSc$_3$) $>$
\textit{a}(InNNi$_3$) $>$ \textit{a}(InCNi$_3$) $>$
\textit{a}(InNCo$_3$) $>$ \textit{a}(ZnNNi$_3$)) as well as the
relationship between bulk modulus and equilibrium volume (i.e.,
$B\sim V^{-1}$)~\cite{Cohen85}. The GGA lattice constants of
InNSc$_3$, ZnNNi$_3$, and InCNi$_3$ are 4.411 \AA~\cite{Maurizio09},
3.77 \AA~\cite{lichong09}, and 3.880 \AA~\cite{Wu20084232},
respectively. The order of shear modulus from low to high is:
\textit{G}(ZnNNi$_3$ and InNNi$_3$) $<$ \textit{G}(InCNi$_3$) $<$
\textit{G}(InNCo$_3$). Pugh~\cite{Pugh54} has proposed that a high
$B/G$ ratio may be associated with better ductility whereas a low
value would correspond to a more brittleness, and the critical value
separating ductile and brittle materials is around 1.75. From the
results of $B/G$ ratio for InNCo$_3$, InNNi$_3$, InNSc$_3$,
ZnNNi$_3$, and InCNi$_3$, it is found that only InNSc$_3$ can be
classified as brittle materials and others may be ductile materials.
Furthermore, InNNi$_3$ seems to be more ductile than InNCo$_3$.

For a cubic crystal, its mechanical stability requires that its
three independent elastic constants should satisfy the following
relations~\cite{Wallace72}:
\begin{equation}\label{eq:9}
(c_{11}-c_{12})>0, c_{11}>0, c_{44}>0, (c_{11}+2c_{12})>0.
\end{equation}
These conditions also lead to a restriction on the magnitude of $B$:
\begin{equation}\label{eq:10}
c_{12}<B<c_{11}.
\end{equation}
The predicted $c_{ij}$ values (see Table~\ref{tab:2}) for InNCo$_3$ and InNNi$_3$ satisfy these conditions,
indicating that cubic antiperovskite-type compounds InNCo$_3$ and
InNNi$_3$ are mechanically stable.

\subsection{\label{sec:estruct}Electronic structures}
In order to understand the different magnetic ground states for InNCo$_3$
and InNNi$_3$, we examined the electronic structures of these two compounds. For the simplicity in discussion, only the GGA results are presented below. The calculated
electronic band structures along the high symmetry directions in the
Brillouin zone are shown in Fig.~\ref{fig:2}. As discussed above,
for InNCo$_3$ the ferromagnetic state is energetically preferable to
the paramagnetic state, and hence it is clearly seen that the
spin-splitting occurs in the bands around the $E_F$. For InNCo$_3$ the profile of
majority spin bands looks roughly similar to the one of
minority spin bands, however only one
band in the majority spin bands crosses the Fermi level and the corresponding band with same dispersion in the minority spin bands is
unoccupied. In addition, two bands in the minority spin bands across the
Fermi level of InNCo$_3$ (see Fig.~\ref{fig:2}(a)). For InNNi$_3$ the
ferromagnetic state is not energetically preferable to the
paramagnetic state, consequently, the majority and minority spin bands are
degenerated, that is to say, no spin-splitting occurs in the
band structure (see Fig.~\ref{fig:2}(b)). It is interesting to note
that the whole feature of majority spin bands of InNNi$_3$ is very
similar to the one of InNCo$_3$. In order to reveal the detailed
character of band structure, the total density of states (DOS) and
the angular-momentum-projected DOS of each atom in InNCo$_3$ and
InNNi$_3$ are presented in Fig.~\ref{fig:3}. For both InNCo$_3$ and
InNNi$_3$, the four bands from -9 eV to -5 eV come mainly from the N
2\emph{p} states and In 5\emph{s} states, and the five bands roughly
from -5 eV to -2.5 eV have significant contribution from
3\emph{d}-$t_{2g}$ states of transition metal atoms (Co/Ni) and the
5\emph{p} state of In atom. For the bands from -2.5 eV to 0 eV (i.e., the Fermi level), they are dominated by the 3\emph{d} states of
transition metal atoms (Co/Ni) and have small contribution from the
2\emph{p} states of N atom. Because the number of 3\emph{d}
electrons in Co is one less than that of Ni, two minority spin
bands composed of the 3\emph{d}-$t_{2g}$ states around the Fermi level  
are unoccupied in InNCo$_3$, while the counterpart in the InNNi$_3$ are occupied. This also results in different behavior of
spin-splitting in InNCo$_3$ and InNNi$_3$. Due to the significant spin-plitting
around the Fermi level in InNCo$_3$, the hybridization between the
Co 3\textit{d} and N 2\textit{p} states in InNCo$_3$ are slightly
weaker than the one between Ni 3\textit{d} and N 2\textit{p} states
in InNNi$_3$(see Fig.~\ref{fig:3}).

The contributions of each kind of atoms to the DOS at the $E_F$ of
InNCo$_3$ and InNNi$_3$ are listed in Table~\ref{tab:3}. The total
DOS at the $E_F$ of InNCo$_3$ is about 2.761 states/eV
per formula unit (f.u.) in the GGA calculations, which is larger than that of InNNi$_3$, and its main contribution comes from Co 3$d$ states which accounts for 87\%. For InNNi$_3$ in the GGA calculations, the contribution of Ni 3\textit{d} states to the
the total DOS at the $E_F$ (i.e., 1.803 states/eV.f.u.) accounts for 72\%. These indicate the 3\textit{d} states of transition metal atoms in InNCo$_3$ and InNNi$_3$ play dominant roles in the total dnesity of states of these compounds. 

In order to understand the bonding nature among the ions in
InNCo$_3$ and InNNi$_3$, we analyzed the charge density contours of
InNCo$_3$ and InNNi$_3$ in the(110) plane, as shown in
Fig.~\ref{fig:4}. From Fig.~\ref{fig:4}, it is found that a certain
amount of charges are accumulated in the intermediate region between
N and Co atoms in InNCo$_3$, and slightly more charges are
accumulated intermediate region between N and Ni atoms in InNNi$_3$.
This gives an evidence for the strong hybridization between N and
transition metal (Co/Ni) atoms, indicating that the N-Co and N-Ni
bondings exhibit strong covalent characteristics and the latter is slightly stronger than the former. The similar
bonding characteristics for Ni-N atoms or Ni-C atoms were also found in
other Ni-based ternary nitrides or carbides
AXNi$_3$~\cite{Wu2009251,Wu20084232,lichong09,Shein10}. Therefore, our results suggest that the magnetic properties of InNNi$_3$ reported in experiment~\cite{Cao093353} are very likely due to the non-stoichiometry effect, which was also found in the cases of AlCNi$_3$ and GaCNi$_3$~\cite{Dong05,Tong06Al,Tong06Ga,Tong07,Sieberer07}.

\section{Conclusions}

In summary, we performed the first-principles calculations to study the
elastic and electronic properties of cubic antiperovskites InNCo$_3$
and InNNi$_3$. Based on the Voigt, Reuss and Hill bounds, the shear,
Young's moduli and Poisson's ratio have also been estimated for the
InNCo$_3$ and InNNi$_3$ polycrystals. The theoretically predicted equilibrium lattice
parameters are in good agreement with the available
experimental data. Our calculations
show that the 3\textit{d} states of transition metal atoms in
InNCo$_3$ and InNNi$_3$ play dominant roles near the Fermi levels.
InNCo$_3$ energetically prefers to the ferromagnetic state. The
magnetic ground state of InNNi$_3$, which is same to other Ni-based
ternary nitrides or carbides with a cubic anti-perovskite structure,
is a stable paramagnetic (non-magnetic) state. This could be understood from
that the hybridization between Ni-3$d$ and N-2$p$ states in
InNNi$_3$ is slightly stronger than the one between Co-3$d$ and
N-2$p$ states in InNCo$_3$ because of the more 3\textit{d} electrons
in Ni.

%%%%%%%%%%%%%%%%%%%%%%%%%%%%%%%%%%%%%%%%%%%%%%%%%%%%%%%%%%%
\section*{Acknowledgments}
The author acknowledges support from National Natural Science
Foundation of China under Grant No. 10674028.
%%%%%%%%%%%%%%%%%%%%%%%%%%%%%%%%%%%%%%%%%%%%%%%%%%%%%%%%%%%

%%%%%%%%%%%%%%%%%%%%%%%%%%%%%%%%%%%%%%%%%%%%%%%%%%%%%%%%%%%
% References
%%%%%%%%%%%%%%%%%%%%%%%%%%%%%%%%%%%%%%%%%%%%%%%%%%%%%%%%%%%
% the natbib package allows both number and author-year (Harvard)
% style referencing;

\bibliographystyle{elsarticle-num}
%\bibliographystyle{apsrev}
%\bibliography{ref}

%%%%%%%%%%%%%%%%%%%%%%%%%%%%%%%%%%%%%%%%%%%%%%%%%%%%%%%%%%%

%%%%%%%%%%%%%%%%%%%%%%%%%%%%%%%%%%%%%%%%%%%%%%%%%%%%%%%%%%%
% Tables and Figures
%%%%%%%%%%%%%%%%%%%%%%%%%%%%%%%%%%%%%%%%%%%%%%%%%%%%%%%%%%%
\clearpage
\newpage

\begin{table}[htbp]
\caption{\label{tab:1} Calculated lattice constants ($a$, in \AA),
bulk modulus ($B$, in GPa), and the first pressure derivative
$B^{\prime}$ of bulk modulus for InNCo$_3$ and InNNi$_3$. $\Delta E_\mathrm{tot}^\mathrm{FM-PM}$ (in eV per formula unit)is the difference between the total energies of ferromagnetic (FM) and paramagnetic (PM) states for InNCo$_3$ and InNNi$_3$. The available
experimental values are also listed. }
\newcommand{\m}{\hphantom{$-$}}
\newcommand{\cc}[1]{\multicolumn{1}{c}{#1}}
\renewcommand{\tabcolsep}{0.25pc} % enlarge column spacing
\renewcommand{\arraystretch}{1.0} % enlarge line spacing
\begin{tabular}{@{}ccc ccc c ccc}
\hline \hline
         & \multicolumn{5}{c}{InNCo$_3$}      &&   \multicolumn{3}{c}{InNNi$_3$}   \\
\cline {2-6} \cline {8-10}
         &   \multicolumn{2}{c}{LDA}    &  \multicolumn{2}{c}{GGA}   & Expt.~\cite{Cao093353}           && LDA   &  GGA  & Expt.~\cite{Cao093353}  \\
         &     PM &  FM &   PM  &  FM  &        &                & PM/FM &   PM/FM     & \\
\hline
 $a$     & 3.744   &  3.753  &  3.835   & 3.855  & 3.8541  & &  3.784   & 3.882   & 3.8445\\
 $B$     & 255.78  & 243.06  &  211.29  & 194.17 &         & &  226.91  & 179.93  &  \\
 $B^{\prime}$ & 4.497 & 4.483 & 5.570   & 5.568  &         & &  4.761  & 4.281    &            \\
 $\Delta E_\mathrm{tot}^\mathrm{FM-PM}$ & - &  -0.0397  &  -  & -0.226 &   &  &0.0& 0.0&\\
\hline \hline
\end{tabular}
\end{table}

\clearpage
\newpage

\begin{table}[htbp]
\caption{\label{tab:2}Calculated elastic constants ($c_{11}$,
$c_{12}$, and $c_{44}$, in GPa), shear modulus ($G$, in GPa),
Young's modulus ($E$, in GPa), and Poisson's ratio ($\nu$) of
InNCo$_3$ and InNNi$_3$. The Voigt shear modulus ($G_V$, in GPa) and
the Reuss shear modulus ($G_R$, in GPa) are also presented. For
comparison, the elastic properties of InNSc$_3$, ZnNNi$_3$, and InCNi$_3$ compounds with a cubic anti-perovskite structure are listed.}
\newcommand{\m}{\hphantom{$-$}}
\newcommand{\cc}[1]{\multicolumn{1}{c}{#1}}
\renewcommand{\tabcolsep}{0.25pc} % enlarge column spacing
\renewcommand{\arraystretch}{1.0} % enlarge line spacing
\begin{tabular}{@{}c c ccc ccc c ccc}
\hline \hline
Compound &  Method  & $c_{11}$ & $c_{12}$ & $c_{44}$ & $G_V$ & $G_R$ & $B$ &$G$  &  $E$   & $\nu$  & $B/G$\\
\hline
InNCo$_3$& LDA &  389.11  &  171.12  & 102.55    & 105.13 & 105.03 & 243.78  &105.08 & 275.63 & 0.311 & 2.320\\
         & GGA &  317.54  &  126.76  &  94.98    & 95.14 &  95.14  &  190.35 &95.14 & 224.67 &  0.286 & 2.001\\
%%%%%%%%%%%%%%%%%%%%%%%%%%%%%%%%%%%%%%%%%%%%%%%%%%%%%%%%%%%
InNNi$_3$& LDA &  356.77  &   164.23  &   69.06  & 80.35 & 78.12 & 228.41 &79.24 & 212.94 & 0.344 &2.895\\
         & GGA &  274.08  &   131.20  &   60.01  & 64.58 & 64.11 & 178.83  &64.35 & 172.37  & 0.339 &2.779\\
%%%%%%%%%%%%%%%%%%%%%%%%%%%%%%%%%%%%%%%%%%%%%%%%%%%%%%%%%%%
%InNSc$_3$& LDA &  252.82 &   59.12  &   91.28  &    93.51 & 93.43 & 123.69 &93.47 &  223.99 &  0.198 & 1.323\\
%         & GGA &  211.98 &   53.34  &   82.63  &    81.30 & 81.28 & 106.22 &81.29 &  194.31 & 0.195 &1.307\\
InNSc$_3$ & GGA~\cite{Maurizio09} &  238.57 &   54.28  &   90.76  &  91.31  & 91.31 & 115.71 &  91.31 &  216.88 & 0.188 &1.267\\
%ZnNNi$_3$& LDA &  423.17 &  153.48  &   56.46  &    87.81 & 73.57 & 243.38 &80.69 &  217.98 &  0.351 & 3.016 \\
%         & GGA &  330.05 &  125.86  &   48.46  &    65.91 & 61.35 & 193.92 &65.63 &  176.96 & 0.348 & 2.955\\
ZnNNi$_3$ & GGA~\cite{lichong09}&   354.28 &   134.01 &   48.06  & 72.89 &   62.05  & 207.43  & 67.47  &   182.61     &  0.353 &3.074    \\
         & GGA~\cite{Shein10} &  364.20 &  124.90   & 32.69  &   67.47 & 46.09  &  204.67   &  56.78 &    155.92      &   0.373 &3.604  \\
InCNi$_3$& LDA~\cite{Wu20084232} &  414.67 &  135.72  &   68.02  &  96.60 & 85.56 & 228.71 &91.08 &  241.22 &  0.324 & 2.511 \\
         & GGA~\cite{Wu20084232} &  344.20 &  106.30  &   62.68  & 85.19 & 77.31 & 185.60 &81.25 &  212.70 & 0.309 & 2.284\\
%MgCNi$_3$ & LDA~\cite{Wu2009251}& 423.71  & 100.04 & 55.21  & 97.86 & 74.97 &207.93  &86.41 & 227.70  & 0.317 & 2.41\\
%          & GGA~\cite{Wu2009251}& 348.68 & 84.03   & 48.88 & 82.26  &  65.37&  172.24 &73.82  & 193.77 & 0.313 &2.33\\
\hline \hline
\end{tabular}
\end{table}

\clearpage
\newpage

\begin{table}[htbp]
\caption{\label{tab:3}Total and partial density of states at the
Fermi level ($N$($E_{F}$), in states/eV.f.u.) for InNCo$_3$ and
InNNi$_3$.}
\newcommand{\m}{\hphantom{$-$}}
\newcommand{\cc}[1]{\multicolumn{1}{c}{#1}}
\renewcommand{\tabcolsep}{0.25pc} % enlarge column spacing
\renewcommand{\arraystretch}{1.0} % enlarge line spacing
\begin{tabular}{@{}c c c ccc}
\hline \hline
         &    &Total density &  \multicolumn{3}{c}{Partial density of states}\\
 \cline {4-6}
         &    & of states      &  In  &    N    &   Co$_3$/Ni$_3$ \\
\hline
InNCo$_3$&LDA &    3.811        &   0.073  &   0.337  &   3.401  \\
         &GGA &    2.761        &   0.144  &   0.169  &   2.448  \\
%%%%%%%%%%%%%%%%%%%%%%%%%%%%%%%%%%%%%%%%%%%%%%%%%%%%%%%%%%%
InNNi$_3$&LDA &    1.580        &   0.160  &   0.212  &   1.208  \\
         &GGA &    1.803        &   0.180  &   0.256  &   1.367  \\
\hline \hline
\end{tabular}
\end{table}

\clearpage
\newpage

\section*{Figure Captions}
\begin{figure}
\begin{center}
\includegraphics*[scale=0.8]{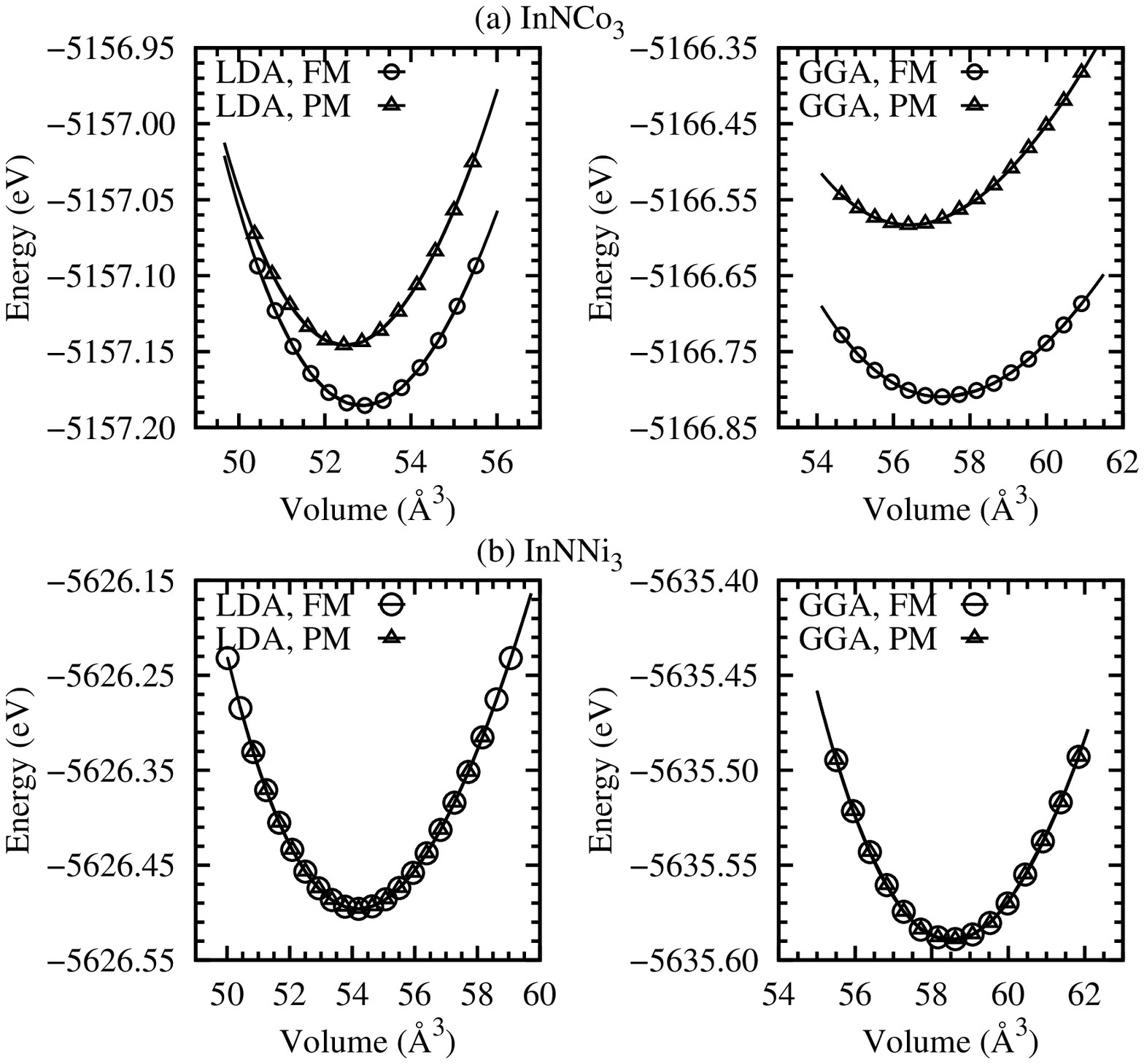}
\caption{\label{fig:1}}
\end{center}
\end{figure}

\begin{figure}
\begin{center}
\includegraphics*[scale=0.8]{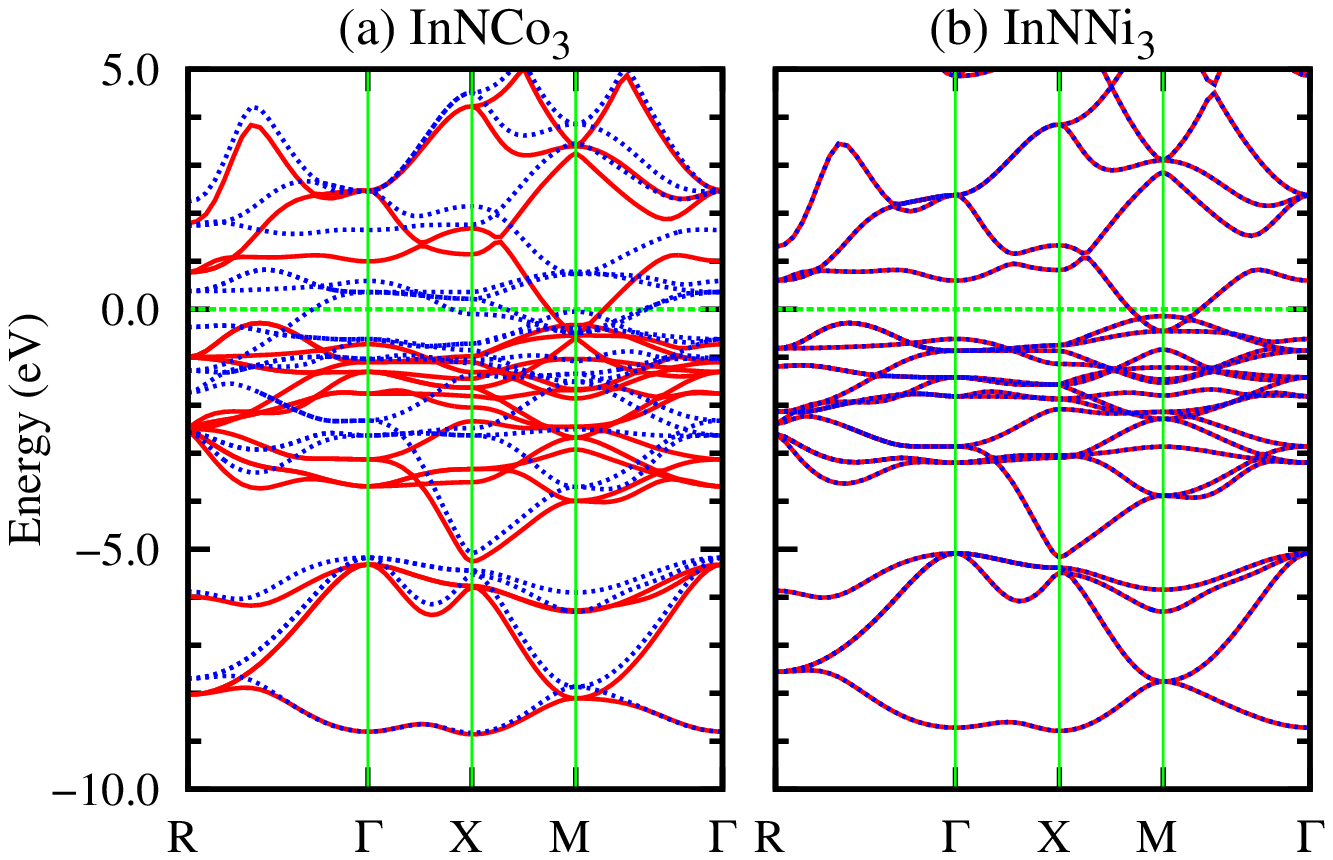}
\caption{\label{fig:2}}
\end{center}
\end{figure}

\begin{figure}
 \begin{center}
   \begin{tabular}{cc}
\includegraphics[scale=0.5,angle=0]{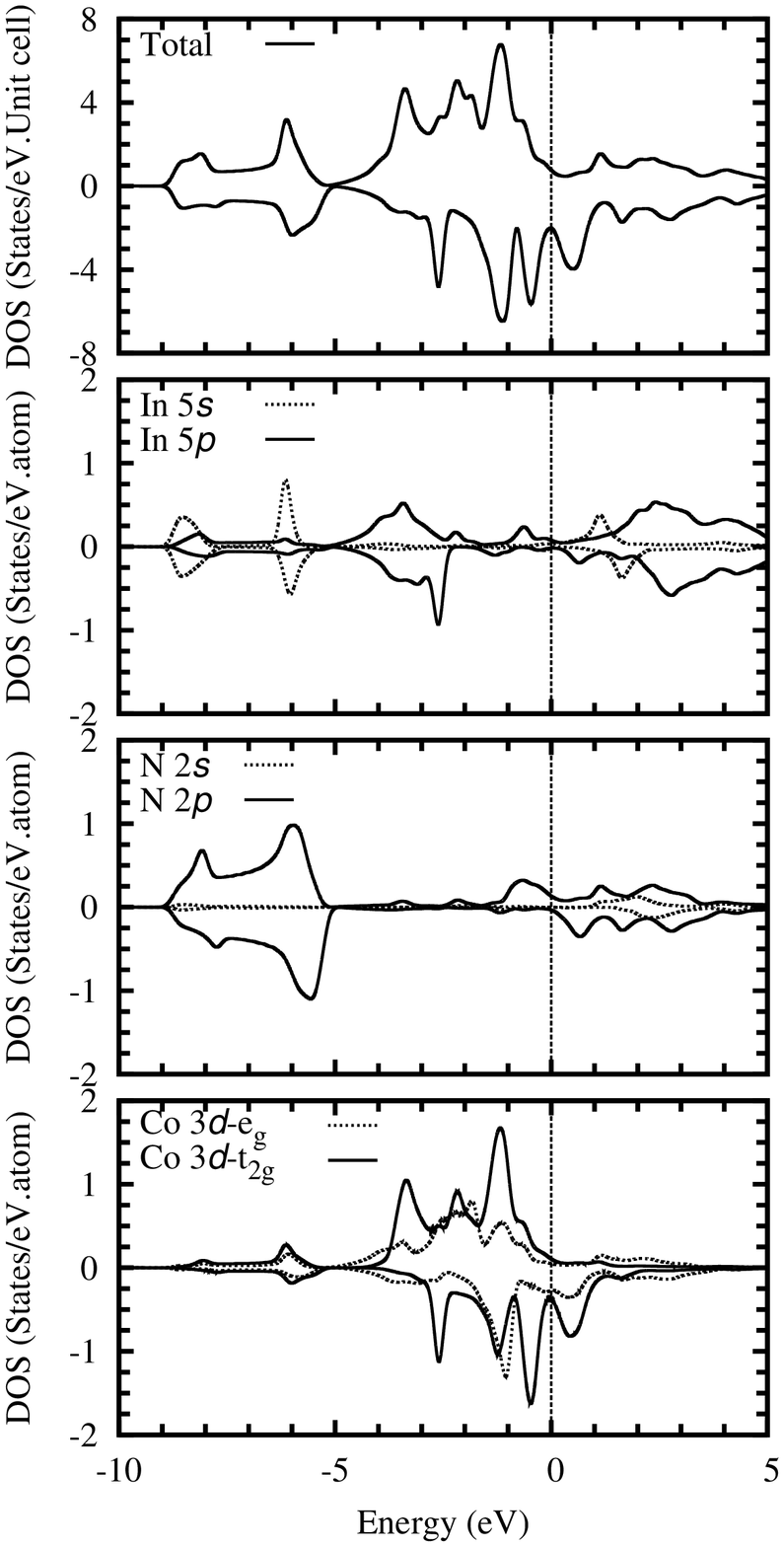} &
\includegraphics[scale=0.5,angle=0]{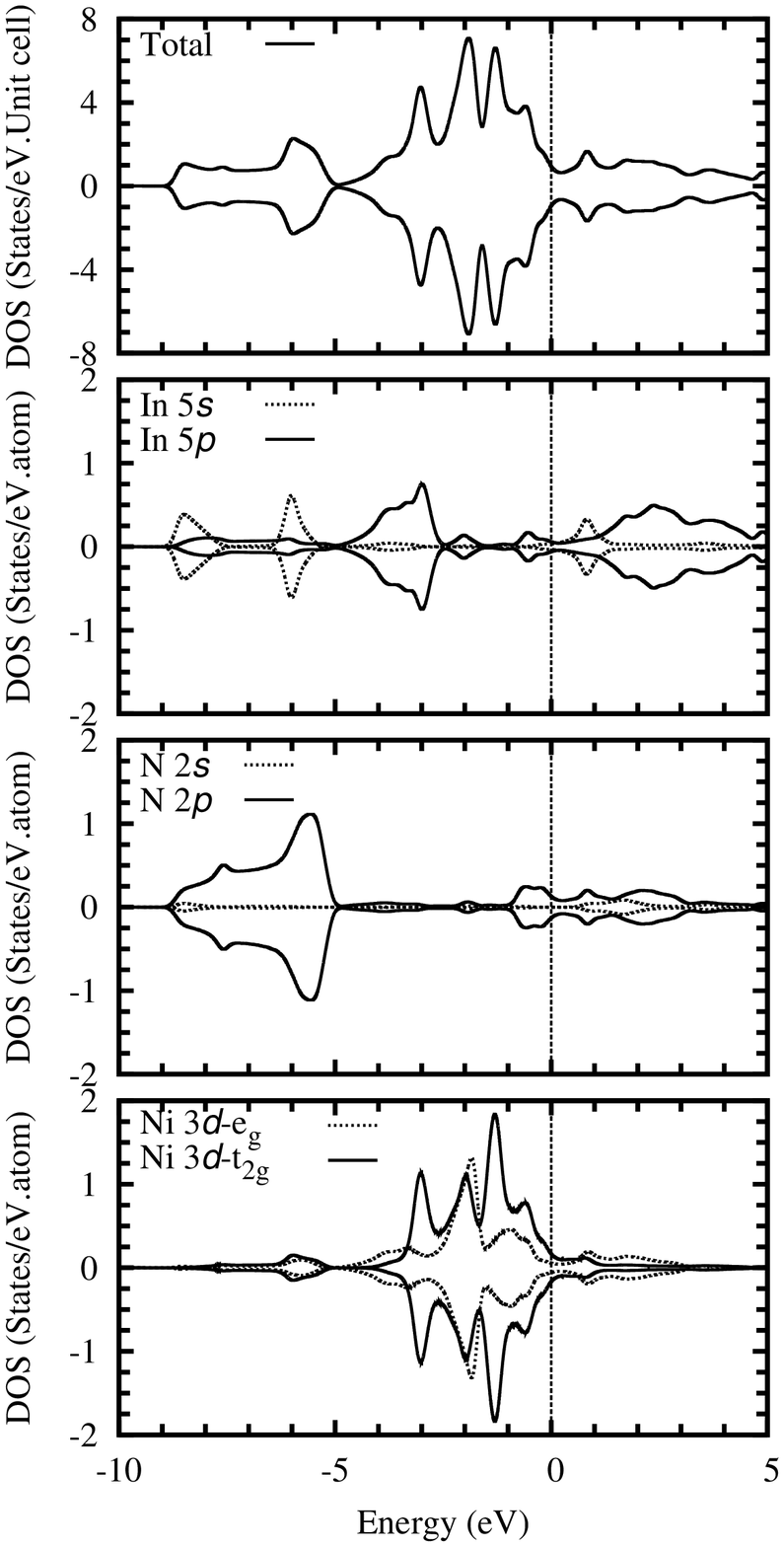} \\
   \end{tabular}
   \caption{}
   \label{fig:3}
 \end{center}
\end{figure}

\begin{figure}
 \begin{center}
   \begin{tabular}{c}
\includegraphics[scale=0.5,angle=-90]{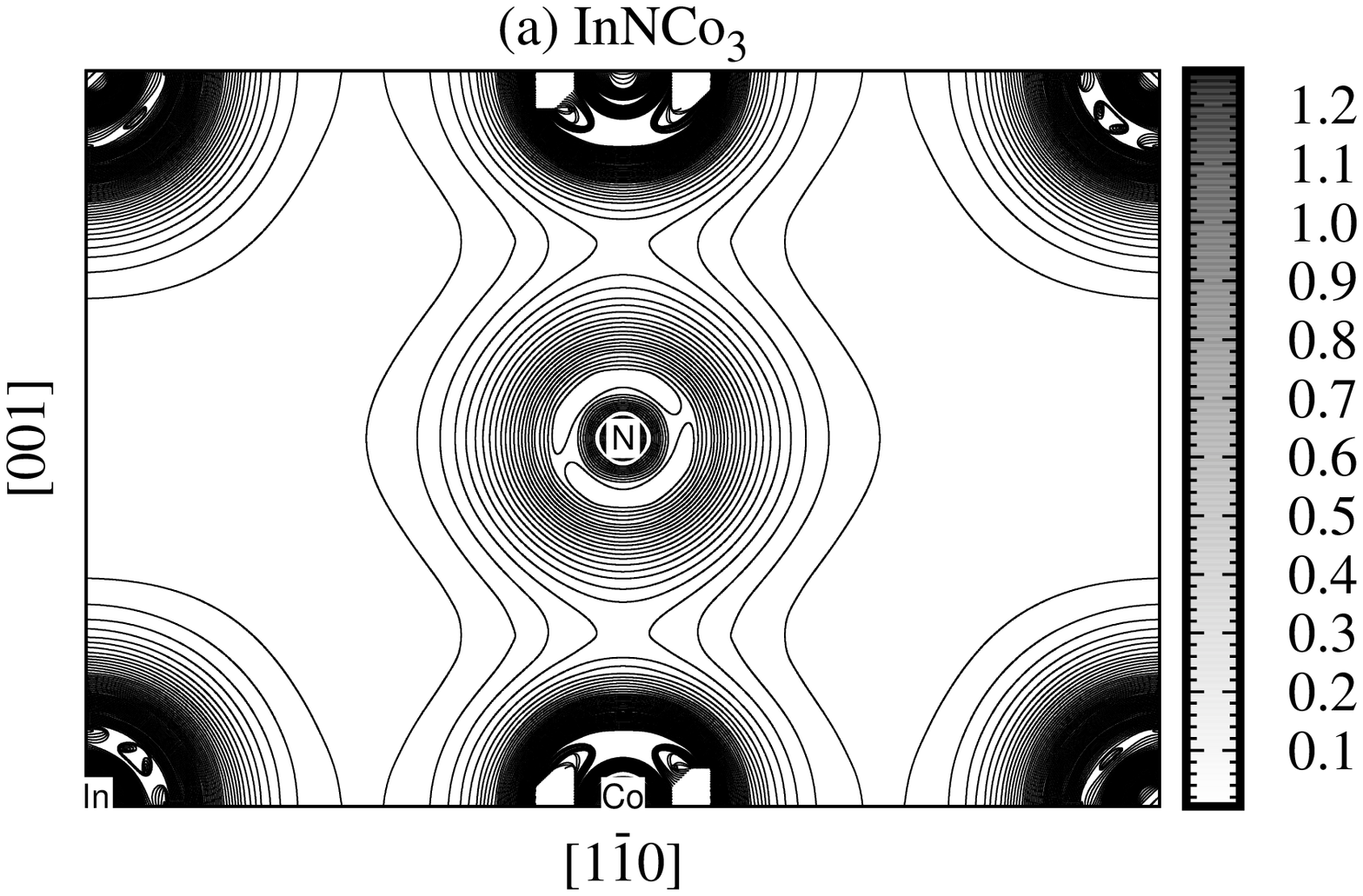} \\
\includegraphics[scale=0.5,angle=-90]{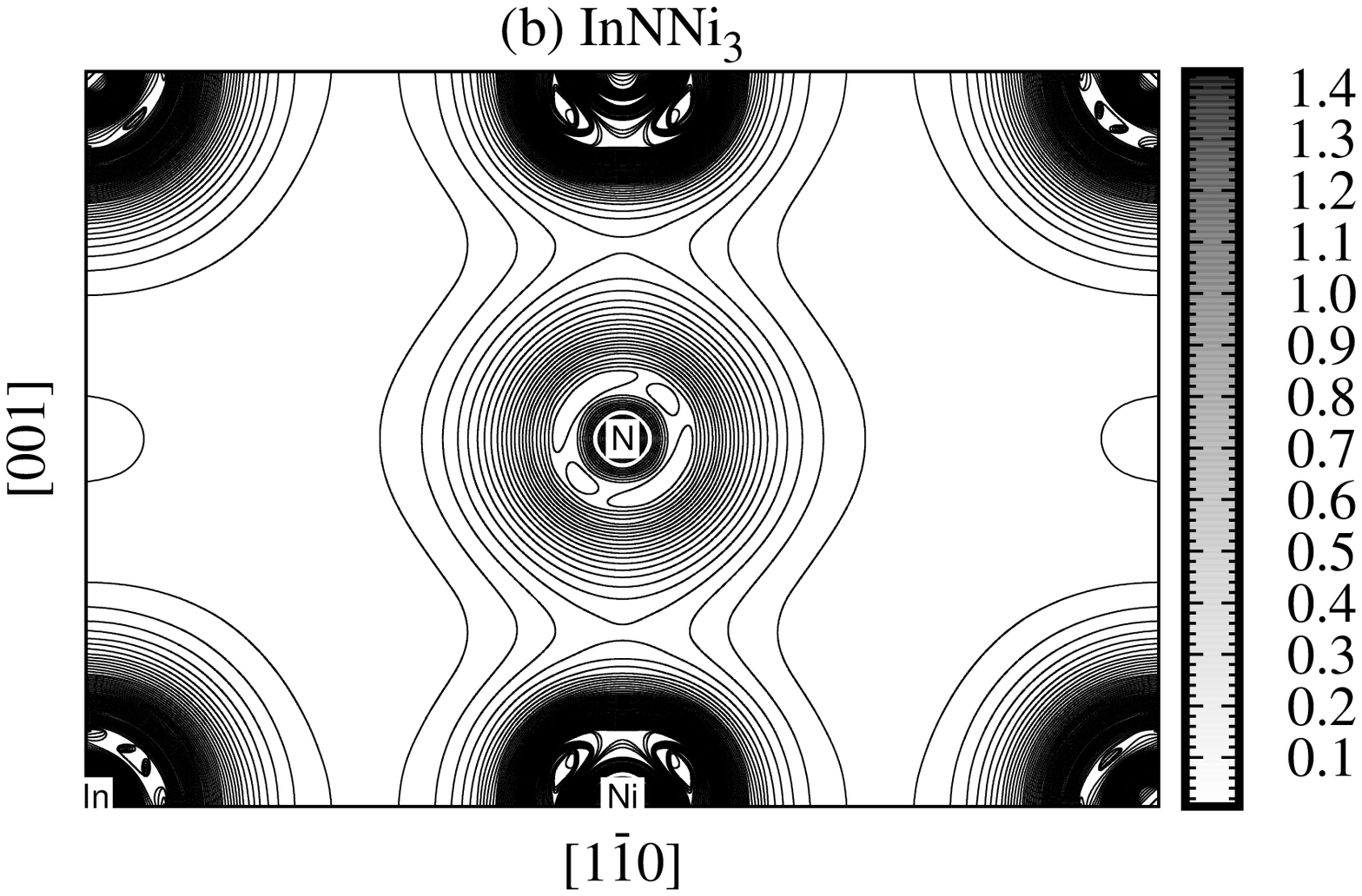} \\
   \end{tabular}
   \caption{}
   \label{fig:4}
 \end{center}
\end{figure}

\begin{itemize}
\item FIG.~\ref{fig:1}: Total energy (in eV per formula unit) versus the atomic volume (in
\AA$^3$) for InNCo$_3$ (upper panel) InNNi$_3$ (lower panel).

\item FIG.~\ref{fig:2}: (Color online) Electronic band structures obtained with GGA
for the majority (red solid line) and minority (blue dashed line) spins of (a)
InNCo$_3$ and (b) InNNi$_3$.

\item FIG.~\ref{fig:3}: Total and partial density of states (DOS) of InNCo$_3$ (left panel) and
InNNi$_3$ (right panel) obtained with the spin-polarized GGA
calculations.

%\item FIG.~\ref{fig:4}: Total and partial density of states (DOS) for InNCo$_3$ obtained with the non-spin-polarized GGA calculations.

\item FIG.~\ref{fig:4}: Charge density contours of the (110) plane for (a) InNCo$_3$ and (b) InNNi$_3$ obtained with the spin-polarized GGA
calculations.

\end{itemize}

\clearpage
\newpage

\clearpage
\newpage

\clearpage
\newpage

\clearpage
\newpage

%\begin{figure}
 %\begin{center}
%\includegraphics[scale=0.8,angle=0]{dos-innco-pm.eps}
%   \caption{}
 %% \end{center}
%\end{figure}

\clearpage
\newpage

\end{document}